\documentclass[apj]{emulateapj}
\usepackage{apjfonts}
\usepackage{times}

\usepackage{amsmath}
\usepackage{graphicx}
\usepackage{natbib}
\usepackage{epstopdf} 
\usepackage{subfigure}
\usepackage{color}

 \font\sevenrm=cmr7 scaled 1000
\shorttitle{}
\shortauthors{Woo et al.}

\newcommand{\mbh}{$M_{\rm BH}$}

\newcommand{\OIII}{O {\sevenrm III}}
\newcommand{\oiii}{[O {\sevenrm III}]}

\newcommand{\Hb}{H$\beta$}
\newcommand{\Ha}{H$\alpha$}
\newcommand{\kms}{km s$^{-1}$}

\begin{document}
\title{A sub-kpc scale binary AGN with double narrow-line regions}
\author{Jong-Hak Woo$^{1,2,6}$}
\author{Hojin Cho$^{1}$}
\author{Bernd Husemann$^{3}$}
\author{S. Komossa$^{4}$}
\author{Daeseong Park$^{1}$}
\author{Vardha N. Bennert$^{5}$}

\affil{
$^1$Astronomy Program, Department of Physics and Astronomy, 
Seoul National University, Seoul, 151-742, Republic of Korea\\
$^2$The Observatories of the Carnegie Institution for Science, 813 Santa Barbara Street, Pasadena, CA 91101, USA\\
$^3$Leibniz-Institut f\"ur Astrophysik Potsdam (AIP), An der Sternwarte 16, 14482 Potsdam, Germany\\
$^4$Max-Planck-Institut f\"ur Radioastronomie, Auf dem H\"ugel 69, 53121, Bonn, Germany\\
$^5$Physics Department, California Polytechnic State University, San Luis Obispo, CA 93407, USA\\
}
\altaffiltext{6}{TJ Park Science Fellow}

\begin{abstract}
We present the kinematic properties of a type-2 QSO, SDSS J132323.33-015941.9 at $z \sim 0.35$,
based on the analysis of Very Large Telescope integral field spectroscopy and 
Hubble Space Telescope (HST) imaging, which suggest that the target is
a binary active galactic nucleus (AGN) with double narrow-line regions. 
The QSO features double-peaked emission lines ([\OIII] and \Hb) 
which can be decomposed into two kinematic components. 
The flux-weighted centroids of the blue and red components are separated by $\sim$0.2\arcsec\ (0.8 kpc in projection) and coincide with the location of the two stellar cores detected in the HST broad-band images, 
implying that both stellar cores host an active black hole.
The line-of-sight velocity of the blue component is comparable to the luminosity-weighted velocity 
of stars in the host galaxy while the red component is redshifted by $\sim$240 \kms,
consistent with typical velocity offsets of two cores in a late stage of a galaxy merger.
If confirmed, the target is one of the rare cases of sub-kpc scale binary AGNs, providing a test-bed 
for understanding the binary AGN population. 
\end{abstract}


\section{INTRODUCTION} \label{section:intro}
Galaxy mergers represent an important stage in the evolution of galaxies. In the course of gas-rich mergers, accretion activity is triggered onto one, or both black holes. As the merger evolves, a bound pair of supermassive black holes (SMBHs) is formed \citep{Begelman1980}, which will 
ultimately coalesce, emitting a giant burst of gravitational wave emission 
(see \citealt{Centrella2010} for a review). Identifying pairs of accreting SMBHs in all stages of galaxy merging is therefore of great interest, hence an active search has been ongoing.
While a few {\em candidate} sub-parsec binary SMBHs 
have been reported based on semi-periodic lightcurves or characteristic structures in radio jets
(e.g., \citealt{Valtonen2011}, see \citealt{Komossa2006} for a review), 
they lack spatially resolved observations, thus still awaiting
confirmation. Spatially resolved spectroscopy is a powerful tool to reveal and confirm the wider pairs of SMBHs, and an intense search is currently ongoing. 
Nevertheless, only a few {\em spatially resolved} close pairs (binary active
galactic nuclei (AGNs)) at $\lesssim$ 1 kpc separation have been confirmed so far through X-ray and radio observations 
\citep{Komossa2003, Rodriguez2006, Fabbiano2011}.

Recently, double-peaked narrow emission lines have been employed to search for binary AGNs,
making use of large spectroscopic data bases, i.e., the Sloan Digital Sky Survey (SDSS)
\citep[e.g.,][]{Zhou2004, Xu2009, Smith2010, Liu2010b, Comerford2012, Fu2012, Ge2012, Liu2013, 
Benitez2013, Barrows13}. 
Apart from binary AGN, however, a number of other mechanisms can potentially produce double-peaked narrow emission lines at separations of a few hundred \kms, including jet-cloud interactions, bi-conical outflows,  special 
narrow-line region (NLR) geometries, or a single AGN illuminating the interstellar media of two galaxies \citep[and references therein]{Xu2009}. 
Therefore, follow-up spatially resolved spectroscopy is essential in selecting true AGN pairs, by identifying
the host galaxy (merger) itself, and the two active cores.
Follow-up observations of double-peakers 
have shown, that only a small fraction (2-10\%) of them likely harbors dual AGNs (e.g., Shen et al. 2011;
Fu et al. 2012; Comerford et al. 2012).
In the radio regime, for example, some sources turned out to be dominated by radio jets
(e.g., Rosario et al. 2010), while others are genuine dual radio-AGN
(e.g., Fu et al. 2011).

Here, we present the discovery of a {\em sub-kpc} scale binary AGN candidate in a type-2 QSO, 
SDSS J132323.33-015941.9 (hereafter J1323), based on Very Large Telescope (VLT) spectroscopy combined 
with Hubble Space Telescope (HST) imaging. 
At the redshift of the QSO, $z=0.35$, 1$\arcsec$ corresponds to $\sim4.9$ kpc, 
assuming  H$_{\rm o}= 70$ \kms\ Mpc$^{-1}$, $\Omega_{\Lambda}=0.7$ and $\Omega_{\rm m}=0.3$.
Very few such sub-kpc systems have been reported so far \citep{Rodriguez2006, Fabbiano2011}.
Their identification is also important for studying the initial
conditions which later define the rapidity of binary coalescence and the amplitude
of gravitational wave recoil.

\begin{figure*}
\includegraphics[width = \textwidth]{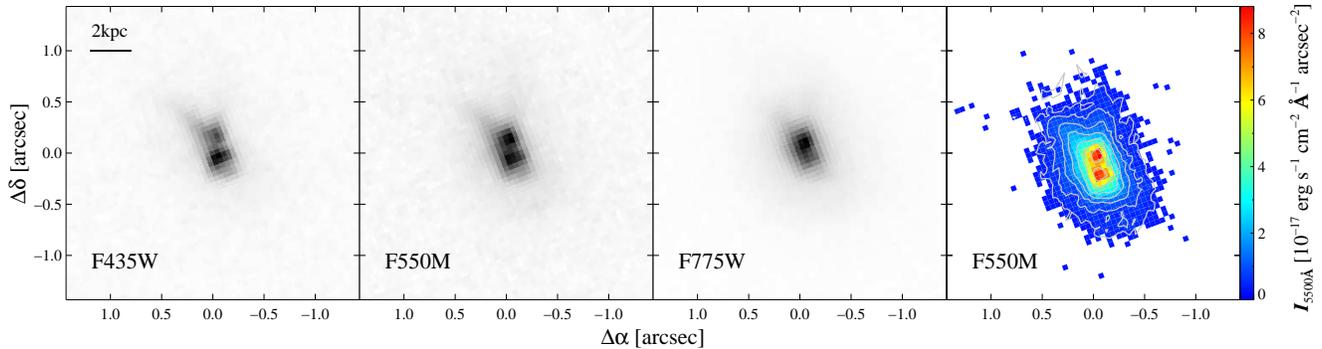}
\caption{
HST images of the host galaxy of type 2 QSO, J1323-0159 observed with three different filers, 
F435W (left), F550M (middle left), and F775W (middle right). 
The flux contours are plotted for the F550M-band image to show the flux contrast (right).}
\end{figure*} 

\section{Observation and Reduction} \label{section:obs}

\subsection{VLT IFU spectroscopy} 

Using the VLT VIMOS instrument \citep{LeFevre:2003}, we obtained integral field unit (IFU) 
spectra of J1323 on February 24 2006, with $\sim$0.7\arcsec\ seeing according to the seeing monitor of the 
observatory. The total exposure time, 5445 sec,  was split into three exposures of 1815\,sec each. We chose the HR grism of VIMOS with a spectral resolution of R$\sim$ 2650, 
covering a spectral range of 5250\AA-7400\AA, which includes the rest-frame [\OIII]$\lambda4959$, 
5007 and \Hb\ lines of the target. The VIMOS field-of-view was set to 13\arcsec$\times$13\arcsec, providing a spatial resolution of 0.33\arcsec\ per spaxel.

The data were reduced by a customized pipeline based on the data reduction software developed for the CALIFA survey \citep{Sanchez:2012, Husemann:2013}. Principle reduction steps include bias subtraction, cosmic ray detection with PyCosmic \citep{Husemann:2012}, fiber tracing, optimal extraction of spectra, wavelength calibration using arc lamp exposures, fiber transmission correction, flux calibration using standard star exposures, and correction for differential atmospheric refraction. VIMOS suffers from instrumental flexures that needs to be properly taken into account during the calibration of each exposure. The flexure effect was characterized in each science exposure by measuring the offset of the fiber traces between the science and continuum exposures 
as a function of wavelength as well as measuring offsets from the expected centroid wavelength of prominent sky lines.

The slightly varying spectral resolutions of the four individual VIMOS spectrographs were characterized across each detector from the width of  emission lines in the arc lamp exposures. 
With an adaptive Gaussian smoothing of the individual spectra we homogenized the spectral resolution to 2\AA\ (FWHM) for each spectrum along the entire wavelength range.

\subsection{HST Images} \label{section:HST}
J1323 was previously observed with the Wide Field Channel of the Advanced Camera for Surveys 
on board HST (Zakamsk et al. 2006), selected as one of the high [\OIII] luminosity type 2 AGNs.
Three filters (F435W, F550M, F775W) were chosen to cover the rest-frame UV, blue and yellow continuum
of the host galaxy, avoiding strong AGN emission lines, i.e.,
[\OIII], \Ha, and \Hb, of the QSO (see Figure 1 in Zakamska et al. 2006).
In these images, we detect a merging pair with two distinct cores, which are separated by
$0.20\arcsec\pm0.01$ with a north-to-east position angle PA = $12.9\arcdeg \pm4.0$, 
based on the flux centroid calculations.
As shown in Figure 1, the flux ratio of the two cores varies depending on the adopted filters,
however, the position of the center of each core and the PA is unchanged.
While the two cores have similar flux in the rest-frame blue-band image, the south core is more 
prominent in the rest-frame UV image. In contrast, the north core is brighter than the south core
in the rest-frame yellow-band image. 
The host galaxy seems to be undergoing a merging process with the two cores separated by a sub-kpc scale in a common envelope.
As a consistency check for the separation and the PA of the two cores,  
we perform the decomposition analysis with 3 Sersic components to fit the two cores and the
envelope in the F550M image, using GALFIT (Peng et al. 2002). Although the image is relatively
shallow to accurately determine the luminosity of each component, we confirm that the separation
($0.22\arcsec\pm0.02$) and the PA ($9.2\arcdeg\pm5.4$) based on the GALFIT analysis is consistent with 
those based on the flux centroid calculation within the uncertainties.

Since the weaker emission lines of the QSO were included in the broad-band photometry,
we quantify the contribution of the narrow-emission lines to the broad-band flux, by convolving
the SDSS spectrum of the QSO with the bandpasses of the aforementioned three filters. 
We find that
the AGN emission line contribution to the broad-band flux is a few per cent, indicating that
the HST images represent the stellar component.

\section{Analyses and Results}\label{section:analyses and results}

\subsection{SDSS spectroscopy} \label{section: stellar kinematics}

To determine the systemic velocity of the host galaxy, we use the SDSS spectrum of J1323
since the VIMOS IFU spectra have low S/N on the continuum with no clear signs of 
stellar absorption lines. In contrast, the SDSS spectrum presents various stellar lines
over the observed spectral range although the level of noise is relatively high. 
We fit the stellar lines at the rest-frame range, 5100-5350\AA, including the Mgb triplet (5172\AA) 
and Fe (5300\AA) lines, using the penalized pixel-fitting software \citep{ppxf} and the 
synthesized stellar templates based on 180 stellar spectra from the MILES library \citep{miles}.
Based on the best-fit template, we derive the line-of-sight (LOS) velocity. 
To account for the uncertainty of the LOS velocity, we perform Monte-Carlo simulations using 
200 mock spectra generated by adding random noise to the SDSS spectrum. 
Then, we fit the stellar lines and  determine the LOS velocity for each spectrum, 
and adopt the 1 $\sigma$ dispersion of the measurements, 24.9 \kms, 
as the uncertainty of the LOS velocity.
Due to the low S/N ratio on the stellar features, the LOS velocity is somewhat uncertain.
Thus, we use the determined systemic velocity as a reference.

Since we used the 3\arcsec\ fiber SDSS spectrum, the derived LOS velocity 
represents the systemic velocity of the luminosity-weighted stellar component in the host galaxy. 
To quantify the luminosity contribution of the north core to the SDSS spectrum, we integrate 
the flux of the north core as well as the flux from the envelope within a 3\arcsec\ aperture, 
using the HST F775W band image. The contribution of the north core is only 10-20\%, indicating 
that the north core is not dominant in the luminosity-weight and that the measured velocity 
represents the systemic velocity of the host galaxy.

\begin{figure*}
\includegraphics[width = .98\textwidth]{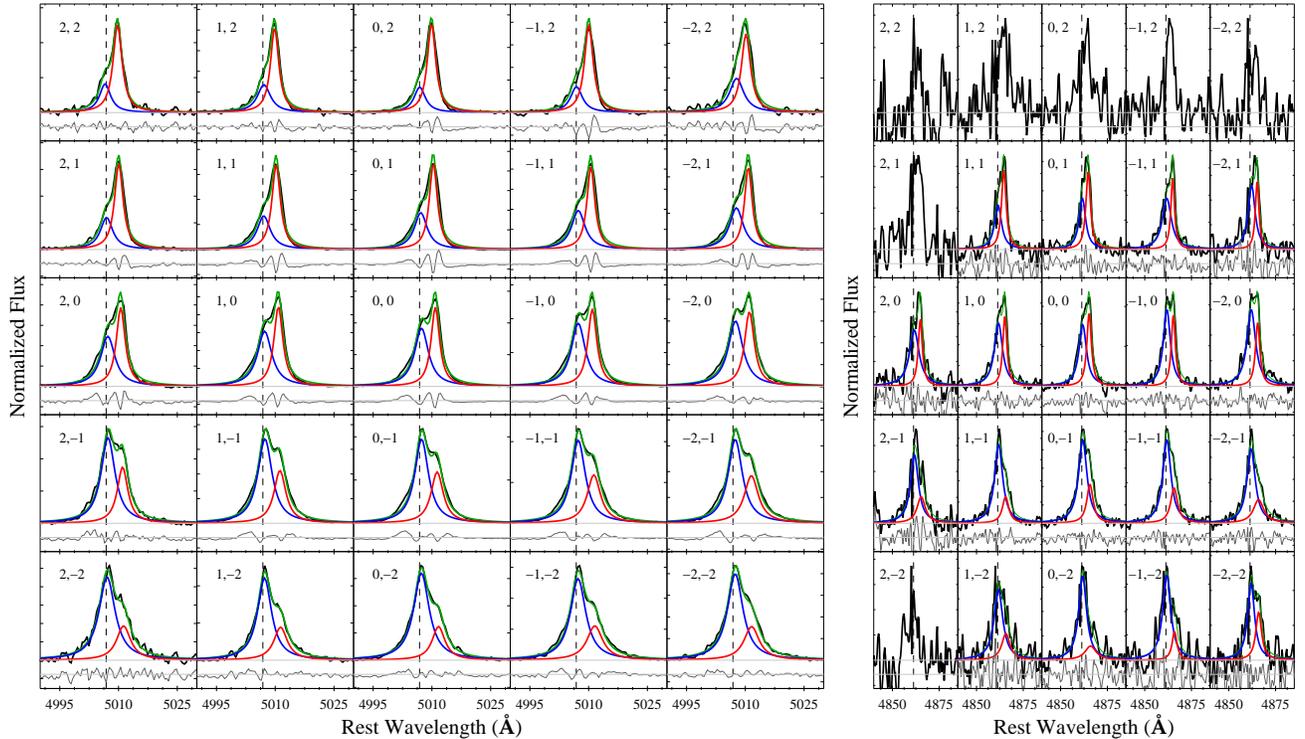}
\caption{
     Examples of the IFU spectra of [\OIII] (left) and \Hb\ (right), 
and their decomposition at the central 5$\times$ 5 spaxels. Black solid lines represent the continuum-subtracted spectra while green lines represent the best-fit model, which is composed of
two kinematic components (blue and red lines, respectively). 
Vertical dashed lines indicate the centers of [\OIII] and \Hb\ lines 
expected from the systemic velocity based on stellar absorption lines. 
Each panel represents a spaxel with a 0.33\arcsec$\times$0.33\arcsec\ scale, centered 
on the spaxel where the [\OIII] line flux peaks.}
\label{fig:allspec}
\vspace{.8cm}
\end{figure*} 

\subsection{Narrow Line Regions} \label{section: narrow line regions}

In Figure 2, we present the observed spectra overplotted with the best-fit model 
for the central $5\times 5$ spaxels, covering a 1.65\arcsec$\times$1.65\arcsec region,
since emission lines are much weaker at outer spaxels.
Here, the center (0,0) is defined as the location of the peak \oiii\ flux, 
corresponding to the location between the north and south cores in the HST images.

As shown in Figure 2, the VLT IFU spectra provide spatial constraints of flux and 
profile changes of the emission lines, \oiii\ and \Hb.
The \oiii\ and \Hb\ lines show a double-peaked profile, particularly at the central spaxels, 
and the flux ratio between the two peaks dramatically changes spatially.
Note that the double-peaked profile was not detected in the SDSS spectrum due to its lower 
spectral resolution. As a consistency check, we combine all spectra from our IFU spaxels, 
then degraded the spectral resolution to match that of the SDSS spectrum. Indeed, 
the combined spectrum shows the same line profile as in the SDSS spectrum.

We kinematically decompose the double-peaked [\OIII]$\lambda5007$ line by fitting the line 
profile with various models, including single, double, and triple Gaussian,
and double Gauss-Hermitian models,  
using the mpfit \citep{mpfit} Levenberg-Marquardt least-square minimization routine.
We find that these Gaussian models do not generate consistent fluxes
between the nearest spaxels, making these models unfavorable since we expect similar fluxes 
in neighboring spaxels due to the seeing.
In contrast, when a pair of Lorentzian profiles was used to fit each emission line, 
we obtain the best results with spatially consistent fluxes. In the fitting process, 
we assume that \oiii$\lambda5007$ and \oiii$\lambda4959$ 
lines have the same redshift and width and that the flux ratio between them is 3:1.
The \Hb\ line was also fitted under the assumption that its LOS velocity coincides with that of \oiii. Although \Hb\ is much weaker than \oiii\ and kinematic decomposition can be done
only for central several spaxels, the \Hb\ line seems to show a similar spatial flux
changes between blue and red components.

\begin{figure}
\includegraphics[width = 0.48\textwidth]{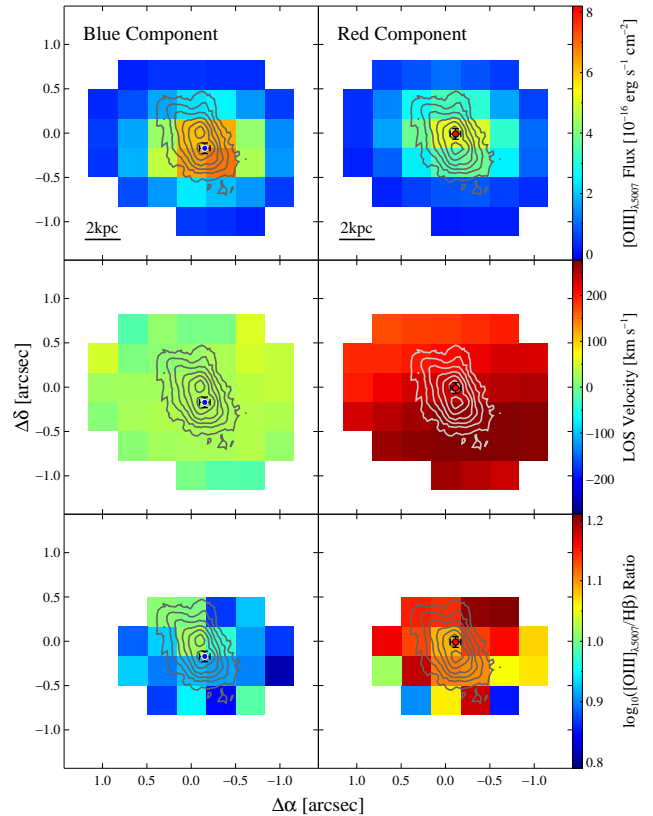}
\caption{
\oiii\ flux map (top), LOS velocity distribution (middle), and \oiii/\Hb\ ratio distribution 
of the blue (left panels) and red components (right panels). 
Each pixel represents a spaxel of 0.33\arcsec$\times$0.33\arcsec. 
Stellar contours measured from the HST F550M-band image are overlapped. 
The flux-weighted centroids of the blue and red components of \oiii\ are denoted with blue and red
circles, respectively, with a 1-$\sigma$ error bar.}
\label{fig:ifudisp}
\end{figure}

In Figure ~\ref{fig:ifudisp}, we present the kinematic analysis, showing the spatial
distributions of the fluxes, LOS velocities, and the [\OIII]/\Hb\ flux ratios for both kinematic components
of \oiii. The [\OIII] flux maps may suggest that both components are extended over kpc scales.
However, we were not able to confirm whether the \oiii\ line gas is extended over kpc scale 
based on our point spread function (PSF) convolution test since the exact PSF at the time of observation is not known.

Although these two components cannot be spatially separated due to the seeing limit ($\sim$0.7\arcsec), 
the IFU data enable us to determine the luminosity-weighted centroid of each component 
as denoted with blue and red points, respectively, in Figure 3. 
The two centroids are separated by $0.17\arcsec\pm0.01$, which is $\sim$0.8 kpc in projection
at the distance of the target, with a PA$=13.9\arcdeg\pm0.5$.
Note that the S/N of [\OIII] in each spaxel is very high, leading to relatively small measurement uncertainties of the flux centroids. 
To better quantify the uncertainty of the separation and the PA, we repeatedly measured the flux centroids 
using various sets of spaxels selected from the [\OIII] map and flux weighting schemes. 
We obtained an average separation 0.17\arcsec\ 
with a 0.01\arcsec\ standard deviation, and a mean PA=13.9\arcdeg\ with a 1.8\arcdeg\ standard deviation. 
We also computed the separation and the PA using the triple Gaussian models in fitting the [\OIII] line,
in order to test the dependence of the line profile modeling on the separation measurement.
While the PA remains consistent with that based on the double Lorenzian models,
the separation becomes slightly larger, presumably due to the uncertainties in modeling the blended wings of [\OIII]. Assuming this discrepancy as the systematic
uncertainty, we conservatively add $\sim$0.06\arcsec\ uncertainty to the measurement error. 
By combining the measurement error (0.01\arcsec), the standard deviation due to the spaxel sampling and weighting
schemes (0.01\arcsec) and the systematic uncertainty in the line profile modeling ($\sim$0.06\arcsec) in quadrature, 
we determine the uncertainty of the separation as 0.06\arcsec. 
Thus, we finalize the measurements as separation = $0.17\arcsec\pm 0.06$ and PA$=13.9\arcdeg\pm1.8$
between the two flux centroids.

Interestingly, the separation and the PA of the two centroids are 
consistent with those of the two stellar cores measured from the HST images.
In Figure 3, we overplot the contours of the host galaxy from the 550M-band image, after shifting the HST image to match the location of 
the [\OIII] flux centroids. Note that since the S/N ratio of the continuum in the IFU spectra is too low to independently 
determine the exact location of the stellar cores, we shift the HST image in order to
match the location of the stellar cores to that of the [\OIII] flux centroids. 
As a consistency check, we compare the flux map of the weak continuum detected in the IFU spectra
with that of the HST image, after convolving the HST image with a typical PSF of the IFU. 
The two flux maps are overlapping each other, 
confirming that the two stellar cores are close to the center of the [\OIII] flux distribution.
The absolute spatial registration between VLT IFU image and HST image is uncertain 
given the current data, hence, future confirmation is required.
However, since the separation and PA of the two components coincide between IFU and HST 
images, it is likely that the flux centroids of the blue and red components of [OIII] 
indeed coincide with the locations of the north and south stellar cores.
Thus, we interpret these results as indication of both stellar cores hosting 
an active black hole with an associated NLR. 

By measuring the flux centroid of the line profile of each \oiii\ component, 
we obtained the velocity offset $\sim$211 \kms\ between blue and red [\OIII] components.
We also measured the LOS velocity of each [\OIII] component with respect to the systemic 
velocity measured from stellar absorption lines. The velocity of the blue component shows 
a small redshift of $30\pm5$ \kms\ with respect to the stellar component while the 
red component is redshifted by $241\pm5$ \kms\ on average with a velocity gradient of $\sim$8 \kms\ kpc$^{-1}$ towards south-west. 
The velocity of the blue component is consistent with the system velocity of the luminosity-weighted
stellar component within the measurement errors ($\sim$25 \kms), 
implying that the active black hole in the south core and the accompanied NLR is close to the
dynamical center of the host galaxy, while the NLR of the active black hole in the north core
is dynamically offset (see Figure 3). 
Assuming two black holes are hosting NLRs, we estimate the mass of each black hole based on the \mbh-stellar velocity dispersions
relationship adopted from McConnell \& Ma (2013; see also Woo et al. 2013), 
using the width of the \oiii\ line as a proxy for the stellar velocity dispersion. 
Although these estimates are highly uncertain, we find that the black hole mass
(log \mbh/M$_{\odot}$) of the blue and red components are 7.6 and 6.7, respectively. 

The [\OIII]/\Hb\ flux ratios of both components are much larger than 3, indicating the ionization source of 
each component is an AGN. The [\OIII]/\Hb\ flux ratio of each component does not vary significantly
over $\pm$1\arcsec\ scale, within which we are able to measure the flux of each \Hb\ components
(see Figure 2). 
In contrast, the [\OIII]/\Hb\ flux ratio between the two components is different by 0.2 dex (58\%).

\section{Discussions and Summary} \label{section:result}

The nature of the double-peaked \oiii\ lines detected with spatially integrated spectra, i.e., SDSS
spectra, is generally ambiguous since it can be explained by various mechanisms:
the projection of two unrelated AGNs, bi-conical outflows, the interaction of a two-sided jet
with NLR clouds, special NLR geometries, a single AGN illuminating the interstellar media 
of two galaxies, or a true binary AGN 
(e.g., Axon et al. 1998; Xu \& Komossa 2009; Crenshaw et al. 2010; Shen et al. 2011).
The uncertainty in interpreting the data arises mostly from the fact, that the (SDSS) 
spectroscopic data do not come with spatially resolved kinematic information
or well-resolved images of host galaxies, as demonstrated by follow-up studies
to overcome the lack of spatial information (e.g., Shen et al. 2011; Fu et al. 2012; 
Liu et al. 2013).

In this paper, we presented a binary AGN candidate based on the spatial 
and kinematic decomposition of the double-peaked [\OIII] line profile using VLT IFU 
spectroscopy and HST imaging. 
Although the spatial resolution of the IFU image is limited to the seeing ($\sim0.7\arcsec$),
the coincidence of the separation and the PA between 
the two stellar cores, with those between the two flux centroids of the [\OIII] kinematic 
components suggests that a binary AGN harbors double NLRs.
As the velocity separation of $\sim$200 \kms\ between blue and red components
is typical of two cores in a galaxy merger, the sub-kpc scale spatial separation may
indicate a binary AGN in a late stage of a galaxy merger.

It is also possible that the double-peaked \oiii\ is due to jet-cloud interaction in a 
single active black hole system. In this case, however, the jet orientation should 
coincide with the PA of the two stellar cores, instead of oriented randomly.
There is a only small chance that the projected jet direction is close to the PA of 
the two stellar cores.
Thus, localized jet-cloud interactions or outflows, and special NLR geometries 
are less viable scenarios for J1323.

Another plausible scenario is that a single AGN photoionizes interstellar media centered on
two stellar cores. 
In this scenario, 
we expect that the fluxes of \oiii\ in two NLRs are different due to the
distance from an ionizing source, covering factor, and gas properties.  
Given the similar strengths of blue and red peaks in the \oiii\ line, 
it is less likely that a {\em single} AGN ionizes {\em two} NLRs, which  
produces a stronger difference in line luminosity under most NLR geometries. 
High resolution radio, optical (e.g., {\it HST}), and X-ray observations are essential 
to confirm that two NLRs are associated with two active nuclei. 
In the X-ray, despite recent improvements in sub-pixel imaging techniques with
{\it Chandra} (e.g., Liu et al. 2013), even marginally resolving two faint cores 
with a $\sim$0.2\arcsec\ separation will be extremely challenging.
The two cores could be easily resolved in the radio, but only if both cores are 
bright radio emitters. In contrast, spatially resolved spectroscopy with HST
would allow us to measure the size, and kinematics structure of the two NLRs, 
confirming the binary nature of J1329.

If confirmed, J1323 is one of the rare cases of sub-kpc scale binary AGNs.
While a handful of binary AGNs have been confirmed with a separation of several kpcs
between two active BHs, only two binary cases show a sub-kpc scale separation, 
i.e., $\sim$10 pc and $\sim$100 pc, respectively 
(Rodriguez et al. 2006; Fabbiano et al. 2011; see also Komossa et al. 2003 for NGC 6240), 
indicating that the discovery of close pairs of BHs in the late stage of galaxy merging is 
observationally difficult, partly due to the challenge of spatially resolving two active cores. 
Thus, confirming the binary nature of J1323
will provide another test-bed for understanding the binary AGN population.

\acknowledgements
This work was supported by the National Research Foundation of Korea (NRF) grant funded
by the Korea government (MEST) (No. 2012-006087 \& 2010-0021558). 
J.H.W. and S.K.
acknowledge the support and hospitality by the Kavli Institute for Theoretical Physics during 
the program, "A Universe of Black Holes". 
This research was supported in part 
by the National Science Foundation under Grant No. NSF PHY11-25915.


\end{document}